\documentclass{svjour2}
\usepackage{graphicx}
\usepackage{rotating}
\usepackage{amssymb}
\usepackage{mathptmx}
\usepackage[numbers]{natbib}
\usepackage{pifont}
\usepackage{marvosym}
\usepackage{dblfloatfix}
\usepackage{subcaption}
\usepackage{subfig}
\usepackage{graphicx}
\makeatletter
\journalname{Journal of Low Temperature Physics}

\begin{document}

\newcommand{\hdblarrow}{H\makebox[0.9ex][l]{$\downdownarrows$}-}
\title{A Comparison of Fundamental Noise in Kinetic Inductance Detectors and Transition Edge Sensors for Millimeter-wave Applications}

\author{A.E. Lowitz \and E.M. Barrentine \and S.R. Golwala \and P.T. Timbie}

\institute{A.E. Lowitz (\Letter) \and E. M. Barrentine \and P.T. Timbie \at 
          Department of Physics, University of Wisconsin - Madison, Madison, WI \\ \email{lowitz@wisc.edu}      
           \and
           S.R. Golwala \at
           	  Department of Physics, California Institute of Technology, Pasadena, CA
}

\date{Received: date / Accepted: date}

\maketitle

\begin{abstract}
Kinetic inductance detectors (KIDs) show promise as a competitive technology for astronomical observations over a wide range of wavelengths.  We are interested in comparing the fundamental limitations to the sensitivity of KIDs with that of transition edge sensors (TESs) at millimeter wavelengths, specifically over the wavelengths required for studies of the Cosmic Microwave Background (CMB). We calculate the total fundamental noise arising from optical and thermal excitations in TESs and KIDs for a variety of bath temperatures and optical loading scenarios for applications at millimeter wavelengths. Special consideration is given to the case of ground-based observations of 100 GHz radiation with a 100 mK bath temperature, conditions consistent with the planned second module of the QUBIC telescope, a CMB instrument \cite{Battistelli2011}. Under these conditions, a titanium nitride KID with optimized critical temperature pays a few percent noise penalty compared to a typical optimized TES.
\end{abstract}

\section{Introduction}
Two types of superconducting, incoherent detectors of millimeter and sub-millimeter radiation have undergone rapid development in the past decade: transition edge sensors and kinetic inductance detectors. TES arrays of 1000s of pixels with micromachined thermal isolation structures have been developed which approach background-limited sensitivity for both ground-based and space-based observations of faint sources such as the CMB.  KID arrays with similar pixel counts have been deployed in instruments such as MUSIC, but they have not yet reached these same sensitivity levels.  However, the simplicity of fabricating and multiplexing KIDs has sparked considerable interest and naturally invites comparison with TES arrays in terms of their ultimate sensitivity.

For an optimized TES, the fundamental noise limits arise from two sources: photon noise from the astrophysical sources or backgrounds under observation, and thermal noise caused by fluctuations in thermal carriers (typically phonons) passing through the weak thermal isolation link from the bolometer's absorbing structure to the thermal bath. For KIDs, photon noise is also present, as is generation-recombination (g-r) noise, caused by fluctuations in the number density of quasiparticles in the device from optical or thermal excitations. It is similar to g-r noise in a photoconductor, caused by fluctuations in the number of charge carriers in a semiconductor. Currently, KIDs are limited by other noise sources, such as two-level system noise in the dielectric substrates, but rapid progress is being made in understanding and reducing these noise contributions \cite{Zmuidzinas2012}.

In this paper we compare the `fundamental noise' limits of KID and TES detectors, particularly as applied to observations of the CMB. We study their performance in both ground-based and space-based optical loading conditions and with bath temperatures below 250 mK, typical of the cryogenic systems used for CMB applications.  We begin by describing the details of the dominant detector noise contributions for KIDs and TESs, as well as our optimization scheme and our assumptions for each detector type.  Finally, we present our comparison results.

\section{Recombination Noise Arising from Optically and Thermally Generated Quasiparticles in KIDs}

The NEP from recombination noise in a superconducting system is \cite{Sergeev1996}:
\begin{equation}NEP_{r} = \frac{2\Delta}{\eta_{pb}}\sqrt{\frac{N_{qp}}{\tau_{qp}}}\end{equation}
where $\Delta$ is the superconducting energy gap.  $\eta_{pb}$ is the efficiency of converting energy into quasiparticles.  $N_{qp}$ is the number of quasiparticles, and $\tau_{qp}$ is the time constant for quasiparticle decay into Cooper pairs.  The general behavior of $\eta_{pb}$ is known from Monte Carlo simulations \cite{Kozorezov2000}: the efficiency approaches 1 when the photon energy is matched to $2\Delta$, and approaches 0.57 when the photon energy is much larger than $2\Delta$.  For our calculations, we have used a simple model: $\eta_{pb} = {2\Delta \over h \nu}$ when $ 1 \geq {2\Delta \over h\nu} \geq 0.57$ and $\eta_{pb} = 0.57$ when ${2\Delta \over h\nu} < 0.57$.

\paragraph{A Simplified Derivation of Optical Quasiparticle Recombination Noise:}

Consider the following simple model of optical quasiparticle creation and decay:
\begin{equation}\frac{dN_{oqp}}{dt} = \frac{P\eta_{pb}}{\Delta}-\frac{N_{oqp}}{\tau_{qp}}\end{equation}
where $N_{oqp}$ is the number of optically-excited quasiparticles and $P$ is the optical power absorbed by the detector.  The first term on the right side of the equation describes optical quasiparticle generation.  The amount of energy per time that contributes to breaking Cooper pairs is $P\eta_{pb}$, and the amount of energy required to excite each quasiparticle is $\Delta$.  The second term on the right side of the equation describes quasiparticle decay.  The rate of quasiparticle decay is $\lambda_{qp} = 1/\tau_{qp}$, and the number of quasiparticles decaying per time scales with the number of quasiparticles present, $N_{oqp}$.  In the steady-state, $dN_{oqp}/ dt = 0$, so $\frac{P\eta_{pb}}{\Delta}=\frac{N_{oqp}}{\tau_{qp}}.$
Simplifying, $N_{qp} = \frac{P \eta_{pb} \tau_{qp}}{\Delta}.$
Using this result in equation 1 gives the following for the optical recombination noise, independent of $\tau_{qp}$:
\begin{equation}NEP_{or} = \sqrt{{4\Delta P/\eta_{pb}}}\end{equation}
which is consistent with the result obtained by Zmuidzinas \cite{Zmuidzinas2012} and others.

\paragraph{Thermally Generated Quasiparticles:}

As in equation 1, the thermally generated recombination noise is
\begin{equation}NEP_{tr} = \frac{2\Delta}{\eta_{pb}}\sqrt{\frac{N_{tqp}}{\tau_{qp}}}.\end{equation}
The number of quasiparticles arising from thermal excitations is \cite{Wilson2001}: 
\begin{equation}N_{tqp} = 2N_0\sqrt{2\pi k_B T_{bath}\Delta}\exp(-\Delta/k_B T_{bath})V,\end{equation}
where $N_0$ is the single spin electron density of states at the Fermi level and $V$ is the volume of the device.  $\tau_{qp}$ can be calculated from $\tau_0$, the material dependent characteristic quasiparticle recombination time, as follows \cite{Kaplan1976}:
\begin{equation}\tau_{qp} = {\tau_0\over\sqrt{\pi}}{N_0(k_B T_c)^{3}\over 2\Delta^{2}}.\end{equation}

\section{Optimization Scheme and Calculation of the Fundamental Noise for a Typical KID}

\paragraph{Optical Recombination Noise:}
To calculate the optical recombination noise we begin with eqn. 3 from above with the approximation that for $T_{bath} \ll T_c$, $ 2\Delta = 3.53k_B T_c .$  The energy gap also determines the pair-breaking efficiency, $\eta_{pb}$ \cite{Kozorezov2000}.

\paragraph{Thermal Recombination Noise:}
Equations 5 and 6 can be used to calculate the thermal noise.  We have calculated the thermal recombination noise for a lumped element titanium nitride (TiN) KID designed to absorb millimeter radiation, with an absorber volume of $\sim 30,000 \mathrm ~\mu m^3$ (for an inductive meander of length $\sim$ 3.7 cm, width $33 \mathrm ~\mu m$, and thickness 25 nm, covering an area of $\sim$ 14 mm$^2$).  NEP dependance on volume is complicated but weak.  Results are qualitatively similar for an order of magnitude range of volumes; for example, an order of magnitude decrease in volume shifts the local minimum and maximum in fig. 1 down in frequency by $\sim$10\%.  TiN offers several advantages over other common superconducting films such as aluminum, including high internal quality factor, long characteristic quasiparticle recombination time, and tunable critical temperature.  For TiN, we have used $N_0 = 3.9\times 10^{10}$ ${\mathrm eV}^{-1} {\mathrm ~\mu m}^{-3}$ and $\tau_0 = 13.7~$ns at 1 K \cite{Gao2012} and scales as $1/T_c^2$, which are reasonable current estimates for these values \cite{Leduc2010}.  

In these calculations of optical and thermal noise, we restrict $T_c$ so that $5\times T_{bath} < T_c < 5 $~K and select the value of $T_c$ which gives the lowest total noise.  We call this the ``optimal $T_c$." Parameters dependent on $T_c$, such as $\Delta$, $\tau_{qp}$, and $\eta_{pb}$, are adjusted accordingly as part of determining the optimal $T_c$. The factor of 5 was chosen in order to be consistent with the approximation that $2\Delta = 3.53k_B T_c$, which is valid only when $T_c$ is well above the bath temperature.

\section{Thermal Noise in TESs}

Neglecting readout and Johnson noise contributions, TES bolometer noise is fundamentally limited by the thermal fluctuation noise occurring across the bolometer thermal weak link.  This noise scales with the temperature of the detector and the temperature-dependent thermal conductance of the weak link, $G(T)$, between the hot absorbing region of the bolometer ($T_{bolo}$) and the cold bath ($T_{bath}$).  For $T_{bolo}\sim T_{bath}$, $NEP_t = \sqrt{4k_B G(T_{bath})T_{bath}^2}$.  However, for CMB optical loading, the TES bolometer experiences a relatively large heating due to the optical and bias loading power, and $T_c \gg T_{bath}$.  In this case, bolometer thermal noise includes an additional term accounting for the thermal gradient across the link as follows \cite{Enss2005}:
\begin{equation}NEP_{t} = \sqrt{4k_B T_{bath}^2G(T_{bath})\frac{n}{2n+1} \frac{\left(\frac{T_{bolo}}{T_{bath}}\right)^{2n+1}-1}{\left(\frac{T_{bolo}}{T_{bath}}\right)^n-1}}\end{equation}
where $n$ is the index of thermal conductivity of the bolometer.  Here we have considered the case of diffuse conduction in the thermal link for the typical case of phonon-phonon scattering across a geometrically long weak link.  Finally, bolometer thermal conductance is defined as: 
\begin{equation}G(T_{bath}) = {\partial P_{tot}\over \partial T_{bath}} = n\kappa T_{bath}^{(n-1)}.\end{equation}
\begin{equation}P_{tot} = \kappa(T_{bolo}^n -T_{bath}^n)\end{equation} is the total power, including optical power and bias power which flows across the bolometer thermal weak link, and $\kappa$ is the temperature-independent thermal conductance coefficient.

\section{Optimization Scheme and Calculation of the Fundamental Noise for a Typical TES}

We begin by assuming an optical loading and a bias power, which determines $P_{tot}$ as follows: $P_{tot} = P_{opt}(n_{bias}+1)$, where $n_{bias}$ is the bias factor.  For a chosen $T_{bath}$, the design requirement $T_{bolo}$ (approximate $T_c$) is then constrained by the choice of $\kappa$ via equation 9.  For each of the loading and $T_{bath}$ scenarios which follow, we choose $\kappa$ to optimize NEP.  We have chosen to use a thermal conductivity index of $n=4$, characteristic of a thermal weak link controlled by phonon-phonon scattering, which is the most common TES bolometer design.  We use a bias factor greater than 1, which is typical to ensure the TES will not be saturated if the optical loading is higher than expected \cite{Niemack2008, Benford2010}.


%
\begin{figure*}
	\begin{minipage}[b]{0.5\linewidth}
	\centering
	\includegraphics[width=.93\textwidth]{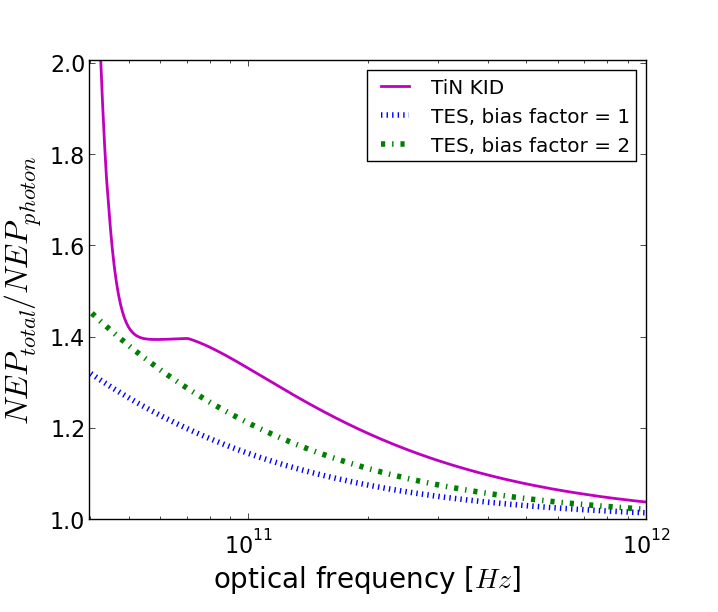}
	\end{minipage}
	\begin{minipage}[b]{0.5\linewidth}
	\centering
	\includegraphics[width=.93\textwidth]{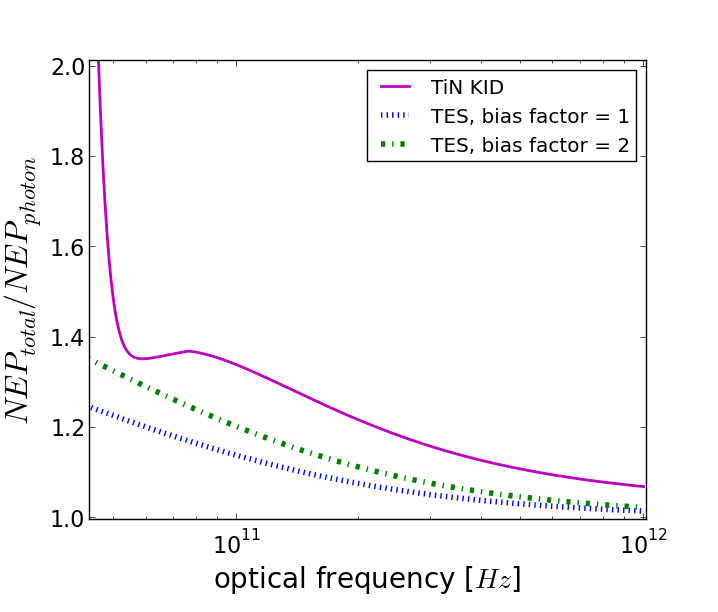}
	\end{minipage}
	\caption{(Color online) Ratio of total NEP to photon NEP as a function of optical frequency for ground-based (left) and space-based (right) observations with a 100 mK bath temperature.}
\end{figure*}	
\begin{figure*}
	\begin{minipage}[b]{0.5\linewidth}
	\centering
	\includegraphics[width=.93\textwidth]{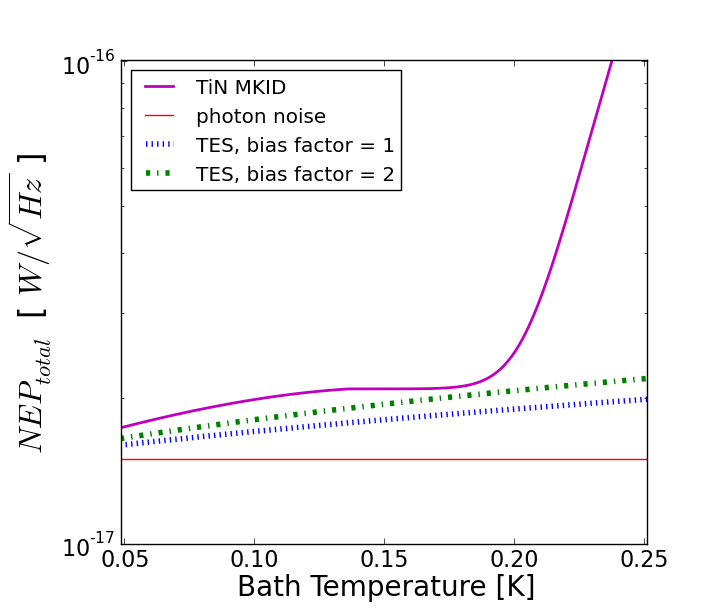}
	\end{minipage}
	\begin{minipage}[b]{0.5\linewidth}
	\centering
	\includegraphics[width=.93\textwidth]{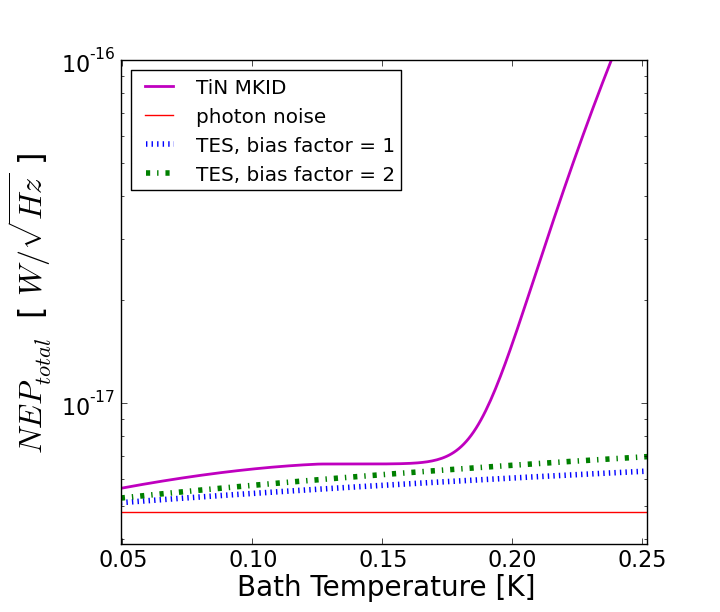}
	\end{minipage}
	\caption{(Color online) Total NEP as a function of bath temperature for ground-based (left) and space-based (right) observations at 100 GHz}
\end{figure*}

\section{Photon Noise and Optical Loading}

We  begin with the usual expression for photon noise, $NEP_{photon} = \sqrt{2Ph\nu(1+mB)},$
where $\nu$ is the center frequency, $B$ is the photon occupation number per mode, and $m = \epsilon \eta$ is the efficiency from emission to detection of one mode. Our results assume optical loading with 30\% bandwidth and optical efficiency of 40\%.  The optical power is $P=h\nu^2\eta B{\Delta\nu \over \nu}$, where $\eta$ is the optical efficiency, and ${\Delta\nu\over\nu}$ is the fractional bandwidth.  In the space-based scenario, a 2.7 K sky temperature with emissivity of 100\% is used.  In the ground-based scenario, a 250 K sky temperature with 4\% emissivity (10 K Rayleigh-Jeans equivalent) is used.  The total noise is simply each noise source, added in quadrature: For KIDs, $NEP_{tr+or+photon} = \sqrt{NEP_{tr}^2 +NEP_{or}^2 + NEP_{photon}^2}$.  For TESs, $NEP_{t+photon} = \sqrt{NEP_{t}^2 + NEP_{photon}^2}$.

\section{Results}

First we consider how the NEP varies with optical frequency for TESs and KIDs.  Figure 1 shows the ratio of the total NEP to the photon NEP as a function of optical frequency with a bath temperature of 100 mK for ground- and space- based scenarios.  Under these conditions, across the range of frequencies considered, the KID has slightly higher noise than the TES with a bias factor of 2.  At 100 GHz, the KID pays a 10\% penalty in noise compared to the bias factor 2 TES in the ground-based case and a 11\% penalty in the space-based case.  
Second, we consider how the NEP varies with bath temperature.  Figure 2 shows the total NEP versus bath temperature with an optical frequency of 100 GHz.  Again, the KID has slightly higher noise under these conditions for the range of bath temperatures considered here.

While KIDs have slightly higher noise in all of these cases, it is important to consider that KIDs can enjoy significant advantages over TESs in fabrication simplicity, multiplexing, focal plane coverage, and tolerance of $T_c$ variation, depending on design.  Ultimately, we believe that KIDs are a competitive technology, especially for high pixel-count arrays.

\begin{acknowledgements}
This work was supported by a NASA Office of the Chief Technologist's Space Technology Research Fellowship.
\end{acknowledgements}


\begin{thebibliography}{99}

\bibitem{Battistelli2011}
E. Battistelli and the QUBIC Collaboration, {\it Astropart. Phys.} \textbf{34}, 705 (2011).

\bibitem{Zmuidzinas2012}
J. Zmuidzinas, {\it Annu. Rev. Condens. Matter Phys.} \textbf{3}, 169, (2012).

\bibitem{Sergeev1996}
A. V. Sergeev and M. YU. Reizer, {\it Intl. J. Mod. Phys. B} \textbf{10}, 635, (1996).

\bibitem{Kozorezov2000}
A.G. Kozorezov, A.F. Volkov, J.K. Wigmore, A. Peacock, A. Poelaert and R. den Hartog, {\it Phys. Rev. B} \textbf{61}, 11807, (2000).

\bibitem{Wilson2001}
C.M. Wilson, L. Frunsio, and D.E. Prober, {\it Phys. Rev. Lett.} \textbf{87}, 067004, (2001).

\bibitem{Kaplan1976}
S.B. Kaplan, C.C. Chi, D.N. Langenberg, J.J. Chang, S. Jafarey, and D.J. Scalapino, {\it Phys. Rev. B} \textbf{14}, 4854, (1976).

\bibitem{Gao2012}
J. Gao et al, {\it Appl. Phys. Lett.} \textbf{101}, 142602, (2012).

\bibitem{Leduc2010}
H.G. Leduc, B. Bumble, P.K. Day, B.H. Eom, J. Gao, S. Golwala, B.A. Mazin, S. McHugh, A. Merrill, D.C. Moore, O. Noroozian, A.D Turner, and J. Zmuidzinas, {\it Appl. Phys. Lett} \textbf{97}, 102509, (2010).

\bibitem{Enss2005}
D. McCammon, "Thermal Equilibrium Calorimeters," {\it Cryogenic Particle Detection,} Springer (2005).

\bibitem{Niemack2008}
M.D. Niemack et al, {\it J. Low Temp. Phys.} \textbf{151}, 690 (2008).

\bibitem{Benford2010}
D.J. Benford et al, {\it Proc. SPIE}, \textbf{7741}, 77411Q (2010).


\end{thebibliography}
\end{document}